\documentclass[preprints,article,accept,moreauthors,pdftex10pt,a4paper]{Definitions/mdpi} 

\firstpage{1}
\pubvolume{xx}
\issuenum{2}
\articlenumber{7}
\pubyear{2019}
\copyrightyear{2019}
\history{Received: date; Accepted: date; Published: date}

\Title{Magnetic Field Vector Structure of NGC6946}

\Author{Kohei Kurahara $^*$ and Hiroyuki Nakanishi}
\AuthorNames{Kohei Kurahara and Hiroyuki Nakanishi}

\address[1]{%
Graduate School of Science and Engineering, Kagoshima University, 1-21-35 Korimoto, Kagoshima 890-0065, Japan; hnakanis@sci.kagoshima-u.ac.jp ~(H.N.)}

\corres{\hangafter=1 \hangindent=1.05em \hspace{-0.82em} Correspondence:  k8791902@kadai.jp}

\abstract{
We studied large-scale magnetic field reversals of a galaxy based on a magnetic vector map of NGC6946. The magnetic vector map was constructed based on the polarization maps in the C and X bands after the determination of the geometrical orientation of a disk with the use of an infrared image and the velocity field, according to the trailing spiral arm assumption. We examined the azimuthal variation of the magnetic vector and found that the magnetic pitch angle changes continually as a function of the azimuthal angle in the inter-arm region. However, the direction of the magnetic field had $180^\circ$ jumps at the azimuthal angles of $20^\circ, 110^\circ, 140^\circ, 220^\circ, 280^\circ$, and $330^\circ$. These reversals seem to be related to the spiral arms since the locations of the jumps are coincident with those of the spiral arms. These six reversals of the magnetic field were seen only in the inner region of NGC6946 whereas four reversals can be identified in the outer region. 
}


\begin{document}
\section{Introduction}
Magnetic field strengths of the order of $\mu$G are observed in many spiral galaxies including the Milky Way~\cite{sofue1986, beck2016, 2018PASJ...70R...2A}. It is known that reversals exist in the magnetic field of the Milky Way based on the observations of pulsars~\cite{2006ApJ...642..868H}. In~addition to the field orientations along the spiral arms, reversals are also found across the galactic plane~\cite{2015ASSL..407..483H}. Recent works show that reversals exist in external galaxies, as~found in IC342~\cite{2015A&A...578A..93B} and in NGC4666~\cite{2019A&A...623A..33S}.\par
We selected NGC6946 to study how these magnetic reversals are generated because it is a nearby grand spiral galaxy located at a heliocentric distance of 7.72 Mpc. Accordingly, this is an ideal sample for the study of the structure of the magnetic field with a good spatial resolution (1' corresponds to 2.2 kpc). Basic parameters for this galaxy are listed in Table~\ref{table:1}. 
Numerous observational studies have been conducted on the magnetic field of NGC6946~\cite{1989A&A...208...32H, 1991A&A...251...15B, 1993A&A...273...45E}. Radio observations at wavelengths of the order of centimeters have shown that this galaxy has a clear magnetic spiral arm, which is bright in synchrotron emission and highly polarized but not associated with stellar spiral arms traced by using optical observations~\cite{1996Natur.379...47B, 2007A&A...470..539B, 2018AN....339..440C}. \citet{2007A&A...470..539B} concluded that the magnetic spiral arms are generated owing to a dynamo~action. 

\begin{table}[H]
\begin{center}
\caption{Basic parameters of~NGC6946.}
\label{table:1}
\begin{tabular}[h]{ccc} \toprule
RA (J2000)& $20^h34^m52.336^s$ & \cite{2007PASJ...59..117K} \\\midrule
Dec (J2000)& $+60^d09^m14.21^s$ & \cite{2007PASJ...59..117K} \\ \midrule
Morphology & SAB & \cite{1992yCat.7137....0D} \\ \midrule
Distances & 7.72 Mpc & \cite{2018AJ....156..105A} \\ \midrule
Position Angle& $242^\circ$ & \cite{2007PASJ...59..117K} \\ \midrule
Inclination & $40^\circ$ & \cite{2007PASJ...59..117K} \\ \bottomrule
\end{tabular}
\end{center}
\end{table}

Additionally, there have been some observational studies on the magnetic field vector. \citet{1997A&A...318..700B} estimated the horizontal and vertical components of the regular magnetic field of the nearby galaxy M51. \citet{2011MNRAS.412.2396F} applied a similar method to M51 and found differences between the magnetic field structures of the disk and the~halo. \par

\begin{figure}[H]
\begin{center}
\includegraphics[width=100mm]{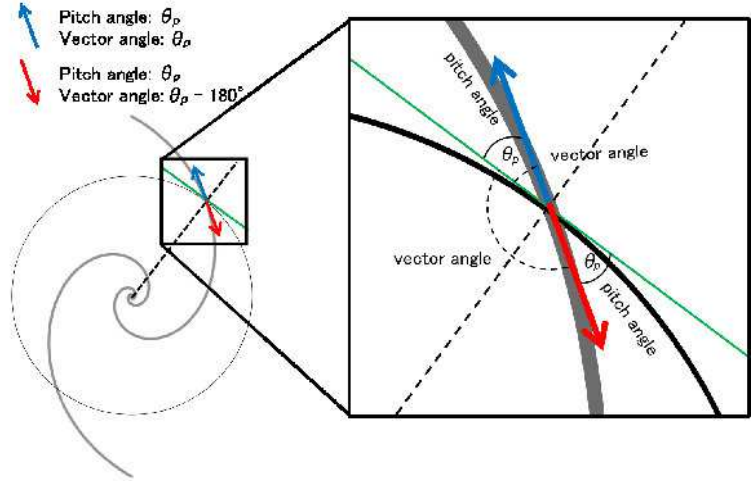}
\caption{Schematic of spiral arms and magnetic field vectors. The~pitch angle $\theta$ is defined as the angle between the tangent of the circle and the magnetic field vector as shown in Figure~\ref{fig:1}. For~a magnetic field directed outwards, the~absolute value of vector angle defined as the same as the pitch angle. On~the other hand, for~a magnetic field directed inwards, it defined as $180^\circ - \theta$. The~sign of vector angle defined as plus and minus for a magnetic field directed outwards and inwards, respectively. These definitions are applied to the magnetic field seen in the face-on view.
}
\label{fig:1}
\end{center}
\end{figure}

Note that in this study, the~pitch and vector angles of the magnetic field are defined as shown in Figure~\ref{fig:1}. The~pitch angle $\theta _p$ is defined as the angle between the tangent to the circle and the magnetic field. The~absolute value of vector angle is defined as the same as the pitch angle for a magnetic field directed outward (counterclockwise in this case) but is defined as $180^\circ - \theta _p$ for a magnetic field directed inward (clockwise in this case). The~sign of vector angle defined as plus and minus for a magnetic field directed outwards and inwards, respectively. Additionally, note that words such as ``orientation'' and ``direction'' are respectively defined as angles with and without the $180^\circ$ ambiguity.  \par

\unskip

%%%%%%%%%%%%%%%%%%%%%%%%%%%%%%%%%%%%%%%%%%%%%%%%%%%%%%%%%%
%%%%%%%%%%%%%%%%%%%%%%%%%%%%%%%%%%%%%%%%%%%%%%%%%%%%%%%%%%
%%%%%%%%%%%%%%%%%%%%%%%%%%%%%%%%%%%%%%%%%%%%%%%%%%%%%%%%%%
\section{Method}
The observed polarization angle $\chi (\lambda ^2)$ can be expressed as a function of the intrinsic polarization angle $\chi_0$, rotation measure (RM), and~the observed wavelength $\lambda$ as follows,

\begin{equation}
\chi (\lambda ^2) = \chi_0 + RM\lambda ^2.
\label{equ:xi} 
\end{equation}

\noindent A vector map of the magnetic field can be derived using the polarization maps generated at two frequencies~\cite{2018_nakanishi}. Based on the definition of the polarization angle at the wavelength $\lambda _i$ as $\chi _i$, the~RM can be estimated from the polarization map of the two frequencies using Equation~(\ref{equ:RM}). 

\begin{equation}
\frac{RM}{\rm rad \, m^{-2}} = 0.81 \int _0 ^{\frac{L}{\rm pc}} (\frac{n_e}{\rm cm^{-3}})(\frac{B_{||}}{\mu {\rm G}}) \, d(\frac{z}{\rm pc}) = \frac{\chi (\lambda _1 ^2) - \chi (\lambda _2 ^2)}{\lambda _1 ^2 - \lambda _2 ^2} ,
\label{equ:RM}
\end{equation}

\noindent where $L$ is the line of sight length, $B_{||}$ is the line-of-sight component of the magnetic field, and $n_e$ is the thermal electron density, respectively. The~sign of the RM is defined to be positive so that the magnetic field vector is directed from the source toward us~\cite{beck2016}. Since polarization maps in lower frequencies than 1.5 GHz have different Faraday depths according to \citet{2017ARA&A..55..111H}, it would be difficult to obtain the RM map correctly in the lower frequency than 1.5 GHz. These problems can be avoided by using C and X band data. The~orientation of the magnetic field can be estimated from a polarization map of the synchrotron radiation at a higher frequency. In~this study, we assumed that the synchrotron emission only arose from a galactic disk, and~that the Faraday rotation occurred mainly within the galactic disk. The~geometrical orientation of the galactic disk was determined based on the assumption that the spiral structure was trailing. The~$180^\circ$ ambiguity of each polarization angle was addressed based on the sign of the RM as explained in the following section, and~the direction of the magnetic field was determined accordingly. Herein, we used an infrared image acquired by the wide-field infrared survey explorer (WISE; \citet{2010AJ....140.1868W}) to trace the spiral arms and the velocity field acquired from the HI nearby galaxy survey (THINGS; \mbox{\citet{2008AJ....136.2563W}}). Based on the THINGS and WISE data we can identify the northwest side of NGC6946 which is nearer to~us. \par

We used data published by \citet{2007A&A...470..539B} and the astronomical image processing system (AIPS) to calculate the polarization intensity and angle using task COMB. Polarization intensity was corrected by subtracting positive bias using task POLCO. Though~polarization maps of NCG6946 shown in our previous paper~\cite{2018_nakanishi} were made with only interferometry data for the simplicity because it is focused on the proposal of the method, the~maps shown in this paper is made with data with single-dish data and the missing large-scale flux is recovered. Basic parameters of the data are listed in Table~\ref{table:2} where the root-mean-square (rms) value of the noise of all Stokes components were calculated in emission-free~regions. 

\begin{table}[H]
\begin{center}
\caption{Relevant data for galaxy~NGC6946.}
\label{table:2}
\begin{tabular}[h]{ccc} \toprule 
\textbf{Observation Band} &\textbf{ C} &\textbf{ X} \\ \midrule
Stokes I rms  & 67~$\mu {\rm  Jy\ beam^{-1}}$&  63~$\mu {\rm Jy\ beam^{-1}}$ \\ \midrule
Stokes Q rms & 17~$\mu {\rm Jy\ beam^{-1}}$& 12~$\mu {\rm Jy\ beam^{-1}}$  \\ \midrule
Stokes U rms & 22~$\mu {\rm Jy\ beam^{-1}}$& 14~$\mu {\rm Jy\ beam^{-1}}$  \\ \midrule
Beam size & $15"$ & $15"$ \\ \bottomrule
\end{tabular}
\end{center}
\end{table}
\unskip

%%%%%%%%%%%%%%%%%%%%%%%%%%%%%%%%%%%%%%%%%%%%%%%%%%%%%%%%%%
%%%%%%%%%%%%%%%%%%%%%%%%%%%%%%%%%%%%%%%%%%%%%%%%%%%%%%%%%%
%%%%%%%%%%%%%%%%%%%%%%%%%%%%%%%%%%%%%%%%%%%%%%%%%%%%%%%%%%
\section{Result}
\begin{figure}[H]
\begin{center}
\includegraphics[width=140mm]{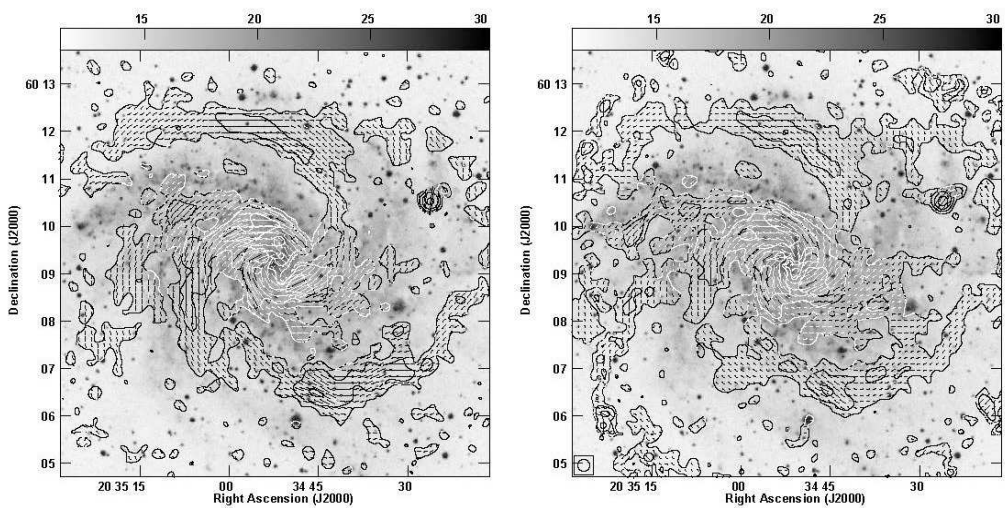}
\caption{Polarized intensity and magnetic field maps of NGC6946 obtained by rotating the observed polarization maps by $90^\circ$ and by overlaying them on a black-and-white image from the 2nd digitized sky survey blue (DSS2 B; \citet{2000ASPC..216..145M}). Left: C-band map. The~contour intervals are (3, 6, 12, 24, 48 and 96) $\times$ 20 $\mu$Jy/beam. Right: X-band map. The~contour intervals are (3, 6, 12, 24, 48 and 96) $\times$ 13~$\mu$Jy/beam.}
\label{fig:2}
\end{center}
\end{figure}

\begin{figure}[H]
\begin{center}
\includegraphics[width=100mm]{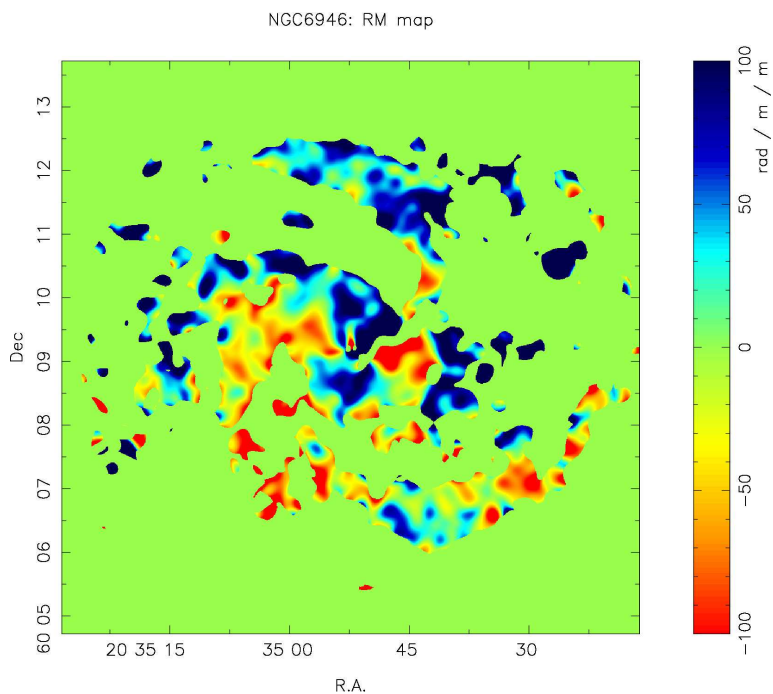}
\caption{The rotation measure (RM) map obtained by applying the Equation~(\ref{equ:RM}) to the C- and X-band polarization maps shown in Figure~\ref{fig:2}. The~foreground RM of $+40\, {\rm rad\ m^{-2}}$ was subtracted based on \citet{2007A&A...470..539B}.}
\label{fig:3}
\end{center}
\end{figure}

We obtained polarization maps in the C and X bands and an RM map, as~shown in Figures~\ref{fig:2} and \ref{fig:3}, respectively. Note that the Faraday rotation effect was not eliminated for the sake of simplicity. Figures~\ref{fig:2} show only polarized intensity values larger than $3\sigma$. 
Even though there is an n$\pi$ ambiguity in the calculation of the difference of the polarization angles between the X and C bands, we considered only the case of n = 0 in the construction of the RM map because the typical |RM| of the galaxy is approximately 100~${\rm rad}\, {\rm m}^{-2}$, which corresponds to the Faraday rotation of $5^\circ$. Because~the foreground RM may cause the magnetic field to change directions, the~foreground RM of +40~${\rm rad\ m^{-2}}$ was subtracted based on \citet{2007A&A...470..539B}.

\unskip

\newpage

\begin{figure}[H]
\begin{center}
\includegraphics[width=75mm]{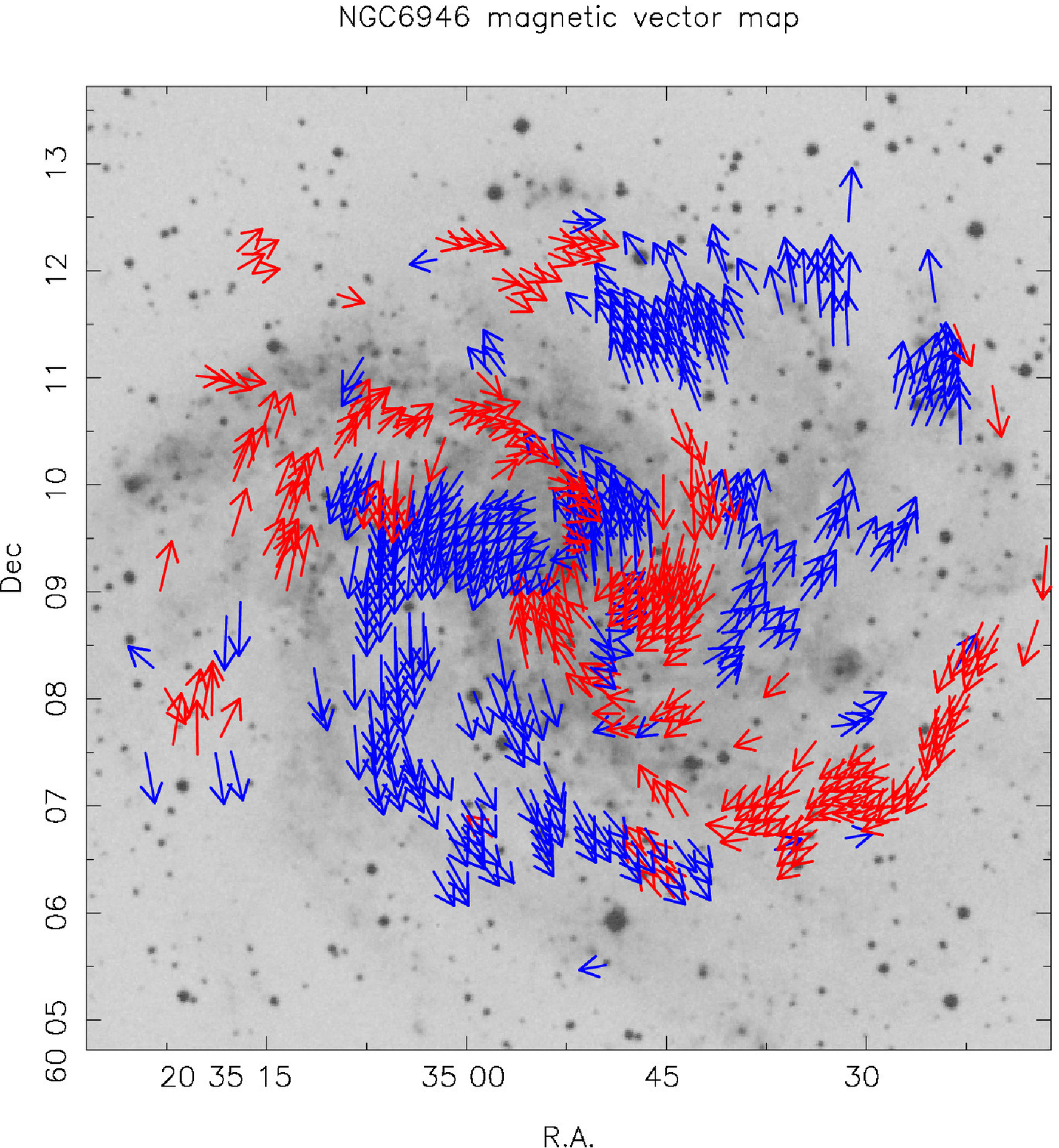}
\includegraphics[width=75mm]{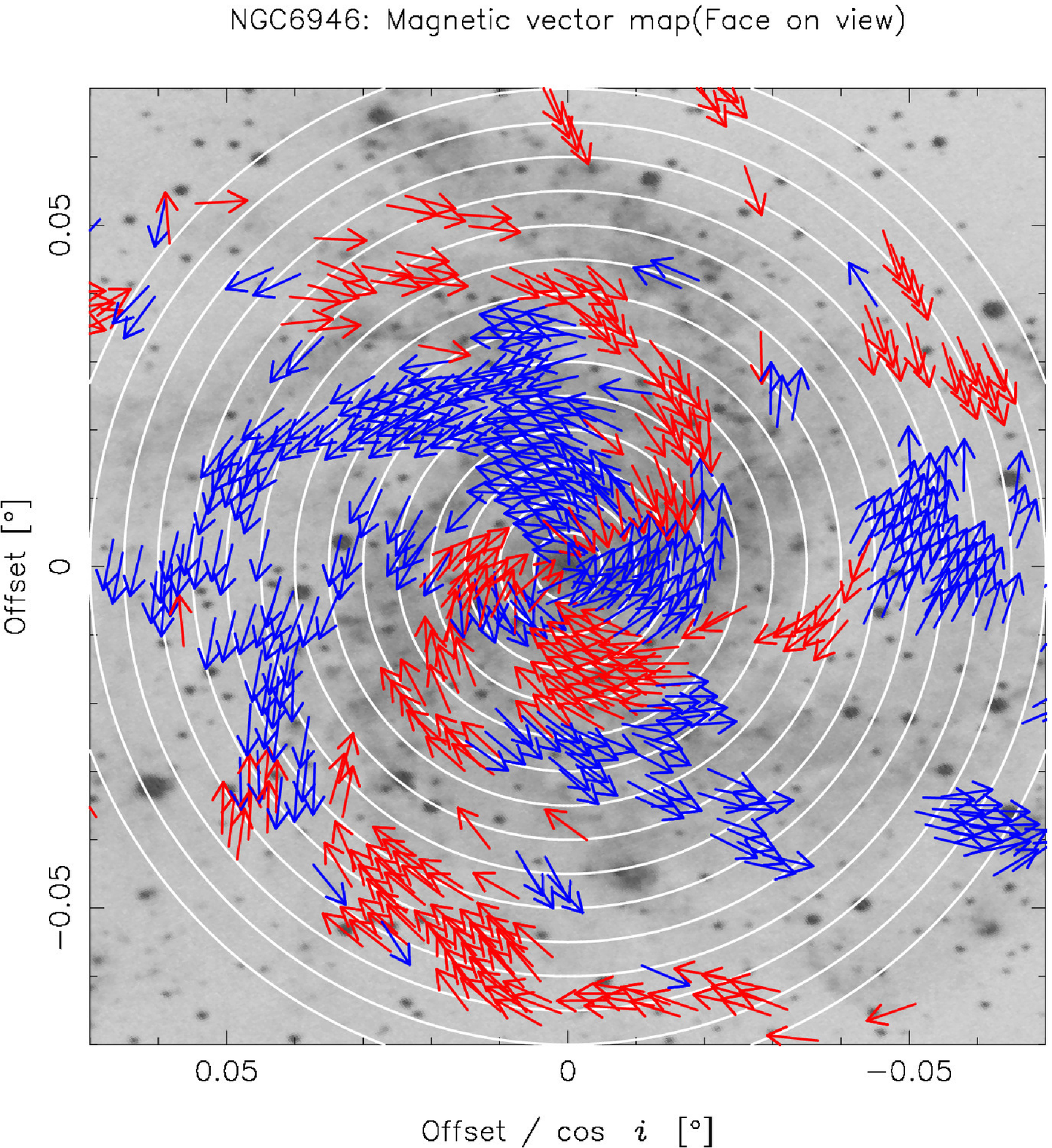}
\caption{Left: Magnetic vector map overlaid on a black-and-white image from DSS2 B. Red and blue arrows denote inward and outward vectors, respectively. Right: Face-on views of the magnetic vector map overlaid on a black-and-white image from DSS2 B. Circles are overlaid at equidistant 680 pc intervals. This map is constructed by rotating the left panel by the position angle and then horizontally enlarging the outcome by a factor of 1/cos {\it i}. Right hand side of the disk is near to~us. }
\label{fig:4}
\end{center}
\end{figure}

Following the method presented by \citet{2018_nakanishi}, we derived a vector map as follows: (1) determining the geometrical orientation of the stellar disk assuming that spiral arms are trailing using the infrared image and the velocity field, (2) obtaining orientations of vectors based on the polarization map at the higher frequency, (3) determining the directions of vectors based on the sign of RMs. As~described in \citet{2018_nakanishi}, data with small absolute value of RM need to be omitted because this method is sensitive to the sign of RM. A~magnetic vector map is, thus, obtained as shown in left panel of Figure~\ref{fig:4}. The~right panel of Figure~\ref{fig:4} shows face-on views of the magnetic vector map. The~face-on view map shown in the left panel was made as follows: (1) rotating the figure on the left panel by $62^\circ$ clockwise so that the major axis is parallel to the y axis, (2) stretching the rotated image by 1/cos {\it i} in the x axis direction to face-on~view.

The magnetic field vectors shown in Figure~\ref{fig:4} are plotted only at points that satisfy the following four criteria: (1) Stokes I is larger than 3$\sigma $ in each band, (2) polarized intensity is larger than 3$\sigma $ in each band, (3) | RM | of each pixel is larger than RM error of that pixel., (4) absolute value of RM is within 200 rad\ m$^{-2}$. The~polarization angle is calculated with $\chi = \frac{1}{2} {\rm arctan} (U / Q)$ using Stokes Q and U. When $\Delta Q$ and $\Delta U$ denote the errors of Stokes Q and U, the~errors of the polarization angle $\Delta \chi$ can be obtained using the error-propagation equation $\Delta \chi = \sqrt{(\frac{\delta \chi }{\delta Q})^2 \Delta Q^2 + (\frac{\delta \chi }{\delta U})^2 \Delta U^2 }$. One pixel has a size equal to $8 \times 8$arcsec square. To~see the magnetic field vector with respect to the spiral arms, it was superimposed on the gray image of the 2nd digitized sky survey blue (DSS2 B; \citet{2000ASPC..216..145M} ).

%%%%%%%%%%%%%%%%%%%%%%%%%%%%%%%%%%%%%%%%%%%%%%%%%%%%%%%%%%
%%%%%%%%%%%%%%%%%%%%%%%%%%%%%%%%%%%%%%%%%%%%%%%%%%%%%%%%%%
%%%%%%%%%%%%%%%%%%%%%%%%%%%%%%%%%%%%%%%%%%%%%%%%%%%%%%%%%%
\section{Discussion}
\unskip
\subsection{Azimuthal Change of the Pitch~Angle}
\begin{figure}[H]
\begin{center}
\includegraphics[width=100mm]{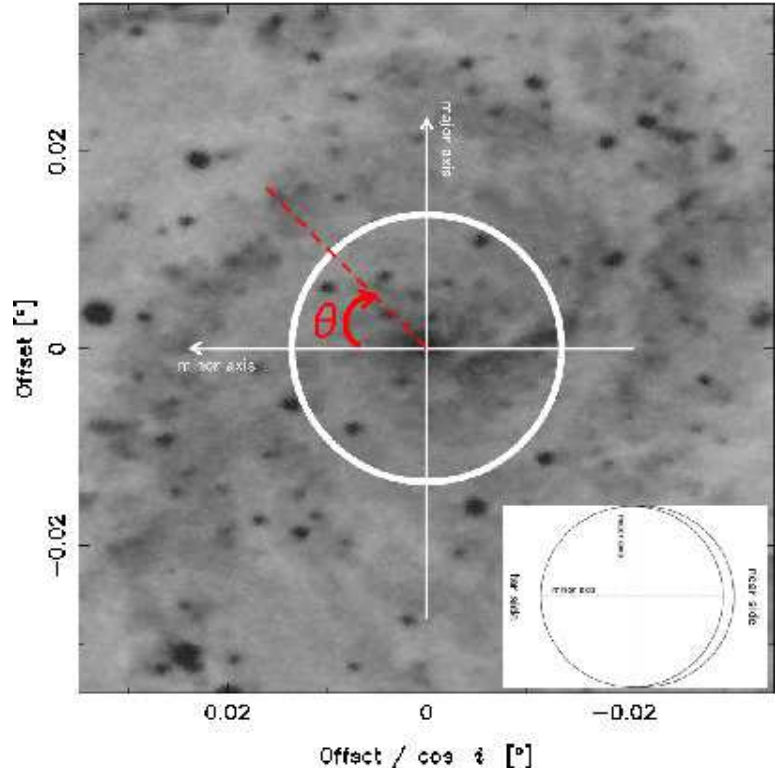}
\caption{Definition of azimuthal angle in Figure~\ref{fig:6}. The~background image is the same as the right panel of Figure~\ref{fig:4}. The~azimuthal angle increases clockwise from the far end of the minor axis. The~radius of the white circle is 1.36 kpc from the center. Right hand side of the disk is near to us.
}
\label{fig:5}
\end{center}
\end{figure}

We examined the azimuthal variations in terms of (1) the pitch angle of magnetic vector and (2) the angle of the magnetic vector by constructing the face-on view map, as~shown in the right panel of Figure~\ref{fig:4}. The~azimuth was defined as an angle measured clockwise from the far end of the minor axis, as~shown in Figure~\ref{fig:5}.

\begin{figure}[H]
\begin{center}
\includegraphics[width=75mm]{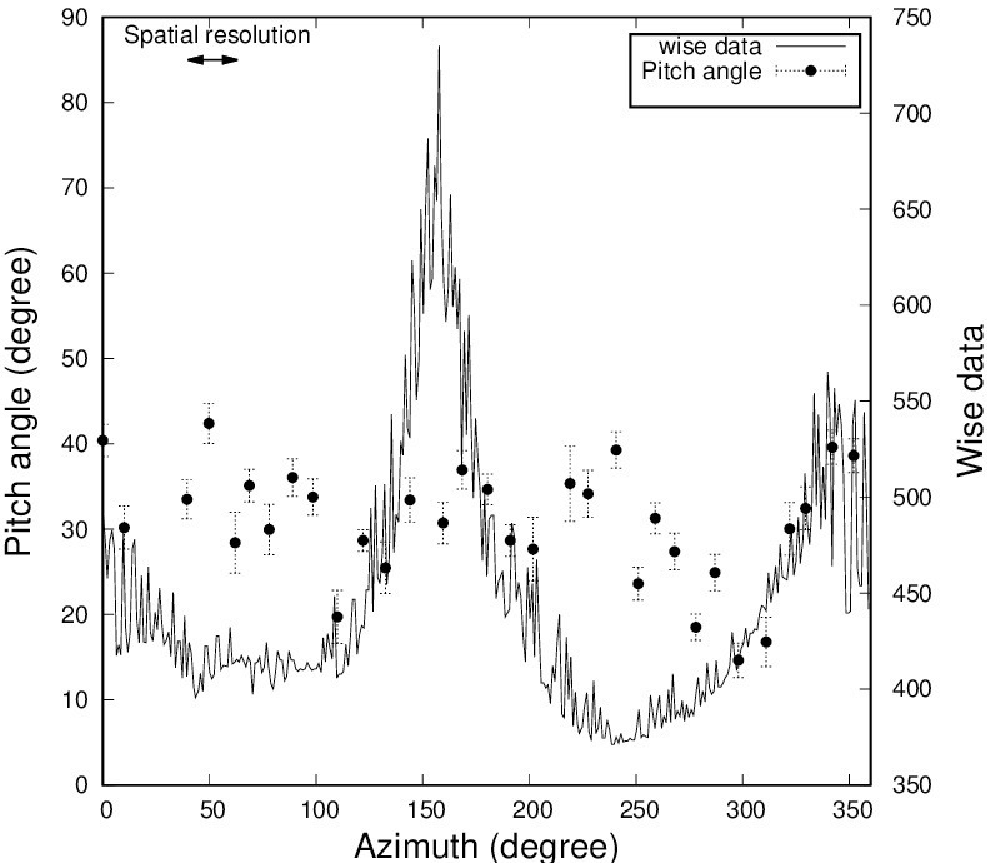}
\includegraphics[width=75mm]{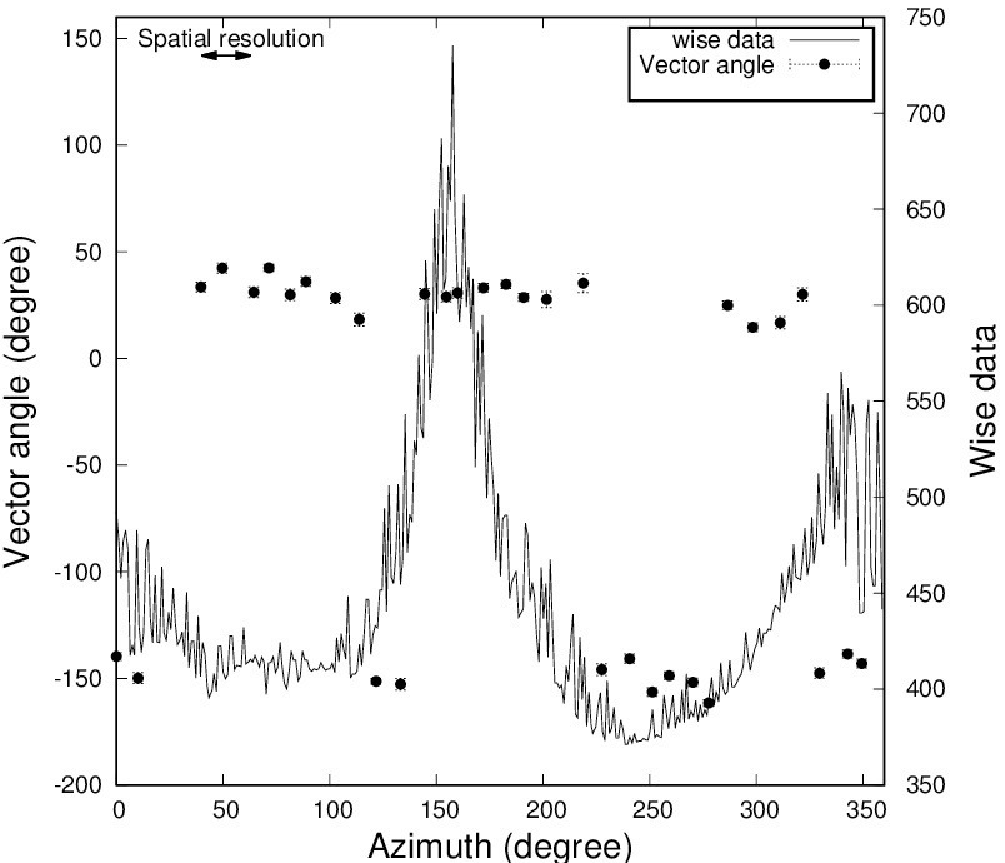}
\caption{Azimuthal variations of magnetic pitch angle (right) and vector angle (left) of NGC6946, respectively. The~solid line denotes the brightness of the WISE image and the peaks correspond to positions of the spiral arms. The~error bars represent the error of the polarization angle of the X band calculated using $\Delta \chi = \sqrt{(\frac{\delta \chi }{\delta Q})^2 \Delta Q^2 + (\frac{\delta \chi }{\delta U})^2 \Delta U^2 }$. Plotted dots are obtained from the annulus between the circles with radii of 0.68 and 2.04 kpc.
}
\label{fig:6}
\end{center}
\end{figure}

The azimuthal variation of the magnetic pitch angle of NGC6946 is shown in left panel of Figure~\ref{fig:6}. Plotted dots are obtained from the annulus between the circles with radii of 0.68 and 2.04 kpc. The~dots in Figure~\ref{fig:6} are averaged values in the azimuth ranges equals to half of the spatial resolution ($\simeq$$12^\circ$) at radius of 1.36 kpc. In~averaging values in different radii, the~azimuthal angle was shifted by $\Delta \theta = \ln{r\over r_0} / {\rm cot}(180 - p)$ with relation to that of radius $r_0=1.02$ kpc using the magnetic pitch angle $p$, which was taken to be $20^\circ$ based on a former work~\cite{2007A&A...470..539B} to eliminate an effect of windings along spiral arms. The~dots in left panel of Figure~\ref{fig:6} represent the pitch angle of the magnetic field (dots and error bars denote mean values and standard deviations, respectively), the~solid lines represent the brightness of the WISE data, and~the peaks correspond to positions of the spiral arms. Looking at the ranges of the inter-arm regions ($40^\circ$--$120^\circ$, $220^\circ$--$300^\circ$) in Figure~\ref{fig:6}, the~azimuthal variation of the magnetic pitch angle changes continually in the inter-arm regions. However, it changes rapidly by approximately $20^\circ$ around azimuthal angles of $280^\circ$ and $110^\circ$, where stellar spiral arms~exist.

\unskip

\subsection{Azimuthal Change of the Vector~Angle}\label{sec4.2}
The right panel of Figure~\ref{fig:6} shows the azimuthal variation of the directions of the magnetic vectors of NGC6946. As~shown in the right panel of Figure~\ref{fig:6}, the~radius range is the same as the left panel, the~dots and solid line indicate the directions of the magnetic field vectors and brightness of the WISE image, respectively. Based on the right panel of Figure~\ref{fig:6}, the~positions of the reversals can be identified at approximately $\theta=20^\circ, 110^\circ, 140^\circ, 220^\circ, 280^\circ$, and $330^\circ$. Along the circle with radius of 1.36 kpc whose angular size is $37"$ the spatial resolution of $15"$ is equivalent to the azimuthal angle of $23^\circ$($=(15/37)\times (180/\pi)$). 
The minimum distance between two adjacent reversals was $30^\circ$ (the case of $\theta=110^\circ$ and $140^\circ$). Therefore, we cannot rule out the possibility that noise makes this jump or could be a local feature, since a jump around $\theta=125^\circ$ is comparable to the resolution. In~the case of the reversal at $\theta=220^\circ$, the vector angle abruptly changes but is constant within the ranges of $\theta=140^\circ$--$220^\circ$ and $\theta=220^\circ$--$280^\circ$, which are much larger than the spatial resolution. Considering that the galaxy moves in the rightward direction in the frame of the right panel of Figure~\ref{fig:6} (rotates clockwise in the frame of Figure~\ref{fig:5}), the~reversals of $\theta=140^\circ$ and $330^\circ$ are coincident with the sides of the upper-stream of the two spiral arms, which peak at approximately $\theta=170^\circ$ and $350^\circ$. This implies that directions of the magnetic vectors change from inward to outward at the upper-stream sides of the spiral arms. The~directions of the magnetic vectors change again after the peaks of the spiral arms. The~other two reversals of $\theta=110^\circ$ and $280^\circ$ are found in the inter-arm region. Since the magnetic spiral arms is found~\cite{2007A&A...470..539B}, these reversals seem to be related to the magnetic spiral~arms. \par

\subsection{Magnetic Field Reversals in the Outer~Region}
\begin{figure}[H]
\begin{center}
\includegraphics[width=150mm]{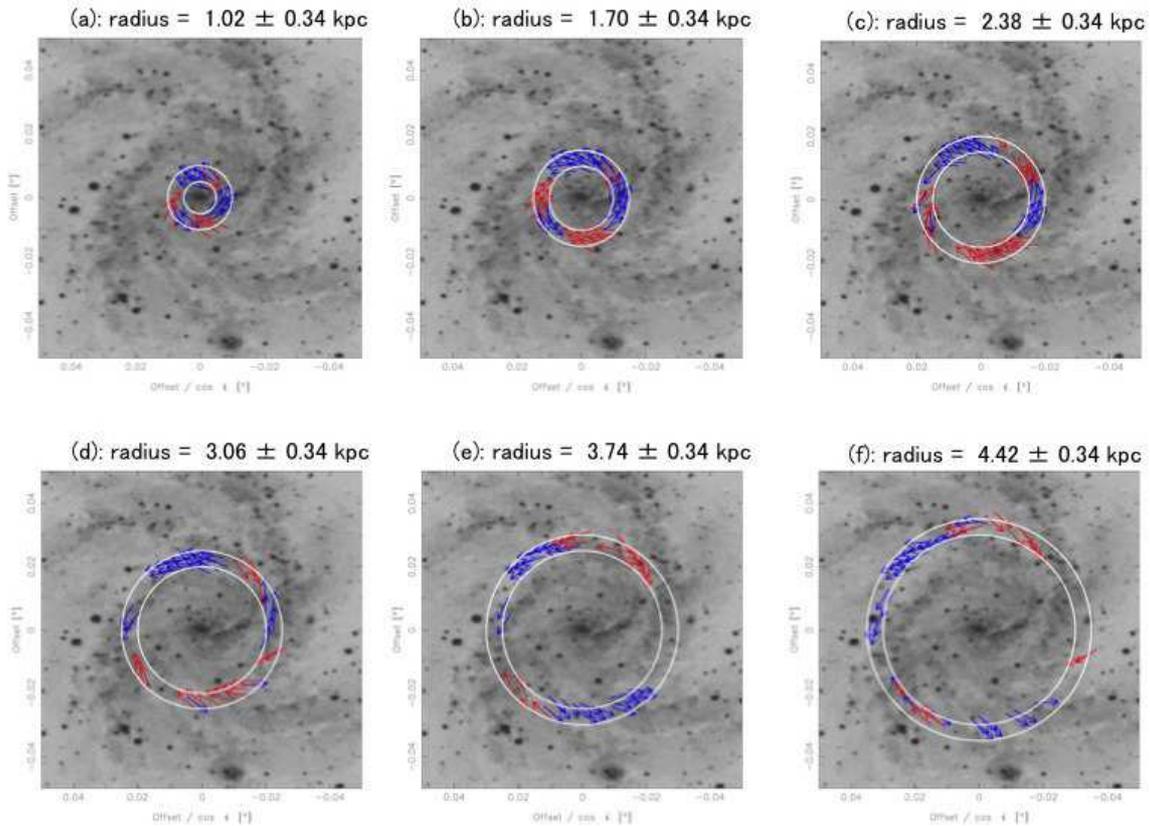}
\caption{Face-on views of the magnetic vector maps of radius ranges of (\textbf{a}) $1.02 \pm 0.34$ kpc, (\textbf{b}) $1.70 \pm 0.34$ kpc, (\textbf{c}) $2.38 \pm 0.34$ kpc, (\textbf{d}) $3.06 \pm 0.34$ kpc, (\textbf{e}) $3.74 \pm 0.34$ kpc and (\textbf{f}) $4.42 \pm 0.34$ kpc overlaid on a black-and-white image from DSS2~B. }
\label{fig:7}
\end{center}
\end{figure}

Next, let us consider if the number of reversals changes with galactocentric distance. Figure~\ref{fig:7} shows face-on views of the magnetic vector map of radius ranges of (a) $1.02 \pm 0.34$ kpc, (b) $1.70 \pm 0.34$~kpc, (c) $2.38 \pm 0.34$ kpc, (d) $3.06 \pm 0.34$ kpc, (e) $3.74 \pm 0.34$ kpc, and (f) $4.42 \pm 0.34$ kpc. As~discussed in the previous Section~\ref{sec4.2}, six reversals were found in (a) and (b) of Figure~\ref{fig:7}.

\begin{figure}[H]
\begin{center}
\includegraphics[width=75mm]{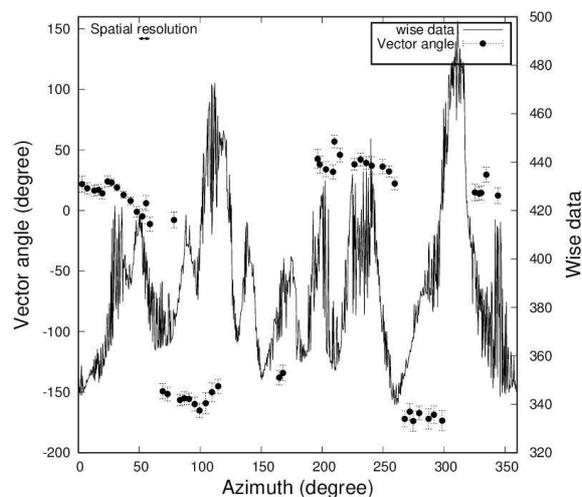}
\caption{Same as the right panel of Figure~\ref{fig:6}, but~for radius range of 3.40--4.76 kpc.
}
\label{fig:8}
\end{center}
\end{figure}

Figure~\ref{fig:8} shows the azimuthal variations of the magnetic vector angles in the the annulus between the circles with radii of 3.40 and 4.76 kpc, which corresponds to panels of (e) and (f) of Figure~\ref{fig:7}. The~dots are averaged values in the azimuth width of $4^\circ$, which is equivalent to the spatial resolution in the case of radius of 4.08 kpc. Figure~\ref{fig:8} shows that there exist four reversals around $\theta = 70^\circ$, $170^\circ$, $260^\circ$, and $310^\circ$ whereas six reversals were found in the inner region. This fact implies that the number of 
reversals of magnetic field might change with the radius. The~mode of m~=~2 identified in the outer regions agrees with \citet{1999A&A...350..423R}, who mentioned that magnetic field structure is superposition of m~=~0 and m~=~2 modes based on the mean-field dynamo theory.

\unskip

%%%%%%%%%%%%%%%%%%%%%%%%%%%%%%%%%%%%%%%%%%%%%%%%%%%%%%%%%%
%%%%%%%%%%%%%%%%%%%%%%%%%%%%%%%%%%%%%%%%%%%%%%%%%%%%%%%%%%
%%%%%%%%%%%%%%%%%%%%%%%%%%%%%%%%%%%%%%%%%%%%%%%%%%%%%%%%%%
\section{Summary}
We applied a method proposed by \citet{2018_nakanishi} to derive a magnetic vector map for NGC6946 using the data of \citet{2007A&A...470..539B} to investigate whether large-scale field reversals exist in external galaxies. First, an~infrared image and a velocity field obtained from WISE~\cite{2010AJ....140.1868W} and THINGS~\cite{2008AJ....136.2563W}, respectively, were used to determine the orientation of the disk. Second, the~vector map was obtained using a polarization map in the X band after resolving the $180^\circ$ ambiguity with the use of the RM which was calculated based on the C and X~bands. \par

The obtained vector map was converted to a face-on view to explore the azimuthal variation of the pitch and vector angles of the magnetic field. As~a result, we found that the direction of the magnetic field changed by $180^\circ$ abruptly, even though the pitch angle changed continually. There are six jumps at $\theta=20^\circ, 110^\circ, 140^\circ, 220^\circ, 280^\circ$, and $330^\circ$, four of which seemed to be related to the position of the stellar arms, and~two seemed to be related to the position of the magnetic arms. We also found that the number of reversals decreases with galactocentric distance $r$ and the outer disk of $r=4.08 \pm 0.68$ kpc has mode of m~=~2, which is consistent with the prediction by the mean-field dynamo theory~\cite{1999A&A...350..423R}.
\vspace{6pt} 

\authorcontributions{Investigation, K.K. and H.N.; Validation, K.K. and H.N.; Supervision, H.N. ; Writing—Original Draft Preparation, K.K.; Writing—review and editing, H.N.
}

\funding{This study has not been specifically funded.}

%%%%%%%%%%%%%%%%%%%%%%%%%%%%%%%%%%%%%%%%%%
\acknowledgments{We are grateful to  Dr Rainer Beck at Max Planck Institute for Radio Astronomy for providing data of this study. We thank to Y. Kudoh for useful~discussions. We also thank to the SOC of “The Power of Faraday Tomography—Towards 3D Mapping of Cosmic Magnetic Fields”. We appreciate useful comments of the reviewers.
}

\conflictsofinterest{The authors declare no conflict of interest.}

\reftitle{References}

%%%%%%%%%%%%%%%%%%%%%%%%%%%%%%%%%%%%%%%%%%
\end{document}